\documentclass[superscriptaddress,prb,letter,aps,cp,amsmath,amssymb,reprint]{revtex4-2}

\usepackage{braket}
\usepackage{lmodern}
\usepackage{graphicx}
\usepackage{dcolumn}
\usepackage{bm}
\usepackage[utf8]{inputenc}
\usepackage[T1]{fontenc}
\usepackage{color}
\usepackage{siunitx}
\usepackage{wasysym}
\usepackage{comment}
\usepackage[inkscapelatex=false]{svg}
\usepackage[colorlinks,citecolor=blue,linkcolor=blue,urlcolor=blue,bookmarks=false,hypertexnames=true, pdfborder={0 0 2},]{hyperref}
\usepackage{xcolor} 
\usepackage[normalem]{ulem}

\usepackage{xspace}

\newcommand{\fref}[2]{%
  \ref{#1}#2\xspace%
}

\graphicspath{{figures/}}

\begin{document}

\newcommand{\fraunhoferIAF}{\affiliation{Fraunhofer Institute for Applied Solid State Physics IAF, Tullastr. 72, 79108 Freiburg, Germany}}
\newcommand{\Takeshi}{\affiliation{National Institutes for Quantum Science and Technology (QST), 1233 Watanuki,
Takasaki, 370-1292, Gunma Japan}}
\newcommand{\Zaitsev}{\affiliation{College of Staten Island (CUNY), Victory Blvd, Staten Island, 10312, New York, USA}}
\newcommand{\Philipp}{\affiliation{School of Science, RMIT University, Melbourne, VIC 3001, Australia}}
\newcommand{\Matthias}{\affiliation{Physikalisches Institut, Universität Heidelberg,
Im Neuenheimer Feld 226, 69120 Heidelberg, Germany}}

\author{Tobias Probst}\thanks{These authors contributed equally to this work.}\fraunhoferIAF
\author{Florian Schall}\thanks{These authors contributed equally to this work.}\fraunhoferIAF
\author{Sebastian Heuft}\fraunhoferIAF
\author{Janina J. Schindler}\fraunhoferIAF
\author{Lukas Lindner}\fraunhoferIAF
\author{Sven Mägdefessel}\fraunhoferIAF
\author{Rüdiger Quay}\fraunhoferIAF
\author{Philipp Reineck}\Philipp
\author{Alexander M. Zaitsev}\Zaitsev
\author{Takeshi Ohshima}\Takeshi
\author{Matthias Weidemüller}\Matthias
\author{Jan Jeske}\email{jan.jeske@iaf.fraunhofer.de}\fraunhoferIAF

\title{Characterizing the nitrogen-vacancy center singlet transition and its phonon sideband for absorption-based room-temperature magnetometry}

\begin{abstract} 
Magnetometry with nitrogen-vacancy (NV) centers in diamond has shown great promise in recent years. In particular, absorption-based magnetometry techniques, employing a cavity to enhance the absorption length, can improve the contrast and sensitivity compared to conventional techniques based on reading out the NV$^-$ triplet fluorescence. The absorption techniques rely on magnetic-field-dependent absorption at the NV$^-$ singlet zero phonon line at \SI{1042}{\nano\metre} and its phonon sideband. In a cavity-enhanced spectroscopy approach, we study pump-laser- and microwave-induced cavity signal changes at room temperature over a spectral range of 680-1050\,nm. Through normalization, we eliminate the cavity-enhancement effect and provide quasi-single-pass values for the absorption and optically detected magnetic resonance (ODMR) contrast. The highest contrast is found at 1042\,nm, but multiple points of high contrast are found at the peaks of the phonon sideband. Additionally, cavity-enhanced ODMR contrasts in the range of 50-80$\,\%$ are presented. We further measure the broadband singlet absorption cross section at room temperature with a novel method through microwave-induced signal changes. This method is insensitive to pump-laser-induced signal changes by other defects and quantifies the room-temperature absorption strength of the singlet transition and its entire phonon sideband. We determine the absorption cross section at 1042\,nm to be $\sigma^{\,\bigstar}_{1042}=(0.89\pm0.14)\cdot 10^{-21}\,\text{m}^2$ or $\sigma^{\,\blacktriangle}_{1042}=(2.9\pm0.5)\cdot 10^{-21}\,\text{m}^2$. depending on the employed \SI{532}{nm} NV$^-$ absorption cross section.
\end{abstract}

\maketitle

\section{Introduction}
\begin{figure*}[htb]
    \centering
    \includegraphics[]{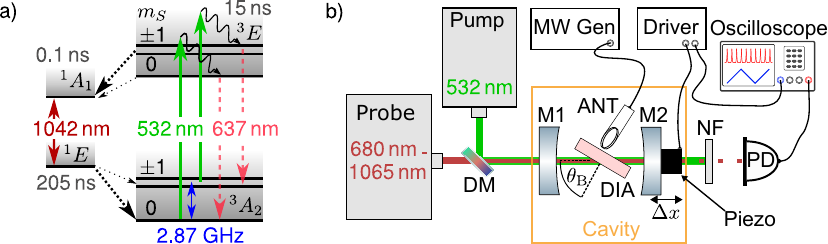}
    \caption{\textbf{NV$^-$ energy level diagram and cavity-enhanced spectroscopy setup.} a) Energy levels of the NV$^-$ center. Optical and microwave transitions are indicated by colors, while phonon-mediated transitions are shown in black. The thickness of the dotted black arrow indicates the probability of the ISC between the triplet and singlet states. The electronic energy levels are broadened due to interactions with phonons (indicated by gray shading). A detailed description of the optical and electronic properties is given in the main text.  b) Experimental setup for cavity-enhanced spectroscopy including a \SI{532}{nm} pump laser and a tunable 680-\SI{1065}{nm} probe laser. Their beams are overlapped with a dichroic mirror (DM) and guided into the cavity, which is designed for the probe laser and created by mirrors M1 and M2. The cavity is scanned with a piezo actuator attached to M2 and contains the diamond (DIA) placed at Brewster's angle ($\theta_{\text{B}}$). Through a microwave loop-antenna (ANT), microwave fields can be applied. A notch filter (NF) blocks the pump light so that only the transmitted probe signal is detected by a photodiode (PD), and is monitored with an oscilloscope.}
    \label{fig:1}
\end{figure*}

Nitrogen-vacancy (NV) centers in diamond have emerged as powerful quantum sensors, especially for measuring magnetic fields \cite{budker2007review, rondin2014review, graham2023fiber}. They can operate at room temperature and under ambient magnetic-field conditions, provide a large dynamic range, inherently enable vector magnetometry, and typically feature a high potential for miniaturization. Conventional ensemble-based NV magnetometry relies on the spin-dependent fluorescence of the NV$^-$ triplet transition, featuring an intensity that varies with the applied magnetic field \cite{rondin2014review}. However, the sensitivity of this approach is currently limited by the weak magnetic-field-induced changes in the fluorescence signal which typically correspond to only a few percent \cite{barry2024sensitive, graham2023fiber}. In particular, the achievable optical contrast is limited by the NV center transition rates and background fluorescence, while the overall signal strength is constrained by limited photon collection efficiency due to the high-refractive index of diamond and geometrical constraints \cite{barry2020review, Xu2019PLdownsides}. 

An alternative approach relies on the magnetic-field-dependent absorption of the NV$^-$ singlet transition, which can be used for magnetometry via transmission measurements of a probe laser \cite{acosta2010singlet}. This approach has been demonstrated for a single-pass configuration, where the signal-to-noise ratio was increased by operating at cryogenic temperatures \cite{acosta2010singlet, acosta2010magnetometry} or using lock-in detection \cite{younesi2025magnetometry}. Alternatively, the signal-to-noise ratio can be enhanced by using an optical cavity to increase the effective absorption path length, thus improving the optical contrast \cite{dumeige2013magnetometry, jensen2014magnetometry, chatzidrosos2017magnetometry}. 

The contrast and signal strength can be further improved by the concept of laser threshold magnetometry (LTM) \cite{jeske2015ltm}. Here, the NV centers are placed in the optical cavity of a laser system, where they provide magnetic-field-dependent gain \cite{jeske2015ltm, jeske2017stimulated, nair2020amp, savvin2021laser, lindner2024ltm, rottstaedt2025ltm} or losses \cite{dumeige2019ltm, nair2021ltm, webb2021ltm, gottesman2024ltm, schall2025ltm, wollenberg2026ltm, lim2026ltm}, strongly dependent on the emission wavelength of the laser system. The loss-based versions of LTM commonly rely on the NV$^-$ singlet absorption. With this technique, contrasts approaching \SI{100}{\percent} were demonstrated, enabling an improved sensitivity limit while still maintaining an unconventionally large dynamic range compared to the state of the art \cite{schall2025ltm}. The advantages of LTM compared to cavity-enhanced and single-pass absorption magnetometry were recently investigated \cite{wollenberg2026ltm, lim2026ltm}. 

All of these absorption-based approaches rely on the zero phonon line (ZPL) of the NV$^-$ singlet transition at \SI{1042}{\nano\meter}. However, this places strong restrictions on the material of the probe laser and on the optical gain medium for LTM. Furthermore, \SI{1042}{\nano\meter} is located close to the spectral cutoff of silicon photodetectors, resulting in a reduced quantum efficiency of these detectors \cite{bludau1974Si_bandgapabsorption}.

Recently, broadband cavity-enhanced absorption magnetometry spectrally separated from the ZPL of the NV$^-$ singlet transition has been demonstrated in the wavelength range from 800-1000\,nm, using the phonon sideband of the singlet transition \cite{schall2025magnetometry}. These sidebands were originally identified by measuring pump-laser-induced transmission changes of a supercontinuum probe laser through an NV-doped diamond at cryogenic temperatures \cite{kehayias2013spectroscopy}. These results were recently reproduced via broadband transient absorption spectroscopy at room temperature \cite{younesi2022broadband}. However, both approaches neglect potential pump-laser-induced contributions from other defects modifying the measurement signals at the respective probe wavelengths. These modifications can occur through stimulated emission or absorption by defects that react directly to the pump-laser light or are formed indirectly in response to processes induced by the pump-laser light. The alteration of these defects could be rooted in a charge-state conversion through either direct pump-induced ionization or indirect capturing of electrons generated by pump-induced ionization processes. Such defects potentially include N$_2$V$^-$ centers \cite{johnson2025nvn}, V$_0$ centers \cite{subedi2021V0stimem}, nickel-related \cite{thiering2021nickel}, and hydrogen-related defects \cite{younesi2025magnetometry}, which all may contribute to the observed pump-induced probe signal changes in the red to near-infrared. 

In this work, we isolate the NV$^-$ effects from these potential other defect contributions via the NV$^-$-specific signal changes induced by a microwave field at \SI{2.87}{\giga\hertz}. We use a cavity-enhanced continuous-wave pump-probe setup to quantify the pump-laser- and microwave-induced transmission changes in the wavelength range from 680-\SI{1065}{nm}. We show an increasing optical contrast toward the ZPL of the singlet transition, reaching maximum values between \SI{50}-\SI{80}{\percent} due to the high-finesse cavity. By normalizing the data with the corresponding cavity enhancement factors, we present the quasi-single-pass signal changes and ODMR contrasts over a broad spectral range. We use microwave-induced cavity signal changes in combination with the Beer-Lambert absorption law and rate model calculations to quantify the absorption strength of the lower NV$^-$ singlet state by determining the room-temperature absorption cross section including its broad phonon sideband. These findings provide new insights into the physics of the NV$^-$ singlet transition and characterize a broad spectral regime for absorption-based room-temperature magnetometry with the potential for significantly improved contrast and sensitivity. 

\section{Experimental setup and methodology}

\begin{figure*}[htb]
    \centering
    {\includegraphics[]{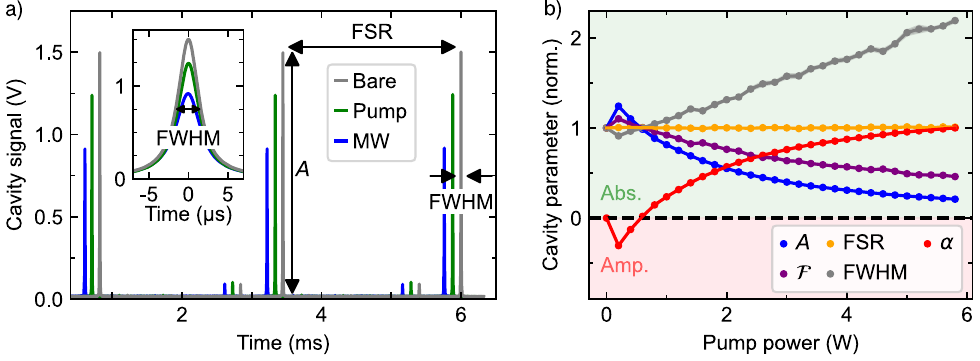}
    }
    \caption{\textbf{Cavity resonances for different experimental configurations and pump power dependence of cavity parameters.} a) Cavity resonances for the bare (gray), pump-induced (green) and microwave-induced while the pump laser is active (blue) cavity signals. Differences in the cavity signal between these three cases are used to calculate the signal changes $\alpha$ and the ODMR contrast, as described in the main text. The cavity resonances are defined by the amplitude $A$, the width (FWHM) and the resonance spacing (FSR), allowing to determine the cavity finesse $\mathcal{F}$. This measurement was performed at a probe wavelength of \SI{900}{nm}, a pump power of \SI{0.85}{W} and a microwave power of \SI{40}{dBm}. b) Pump power dependence of the cavity parameters $A$ (blue), FSR (orange), FWHM (gray) and $\mathcal{F}$ (purple) which are normalized to the values for the bare cavity signal (\SI{0}{W} pump power). From these cavity parameters the corresponding pump-induced signal changes $\alpha$ (red) of the probe light at \SI{900}{nm} are determined, normalized to their maximum value. At this wavelength, both amplification ($\alpha<0$, red shading) and absorption ($\alpha>0$, green shading) are observed depending on the pump power (see main text).}
    \label{fig:2}
\end{figure*}

In our experiments, the NV$^-$ triplet transition ({}$^3A_2\rightarrow{}^3E$) is excited off-resonantly with a \SI{532}{nm} pump laser, as depicted in Fig.$\,$\fref{fig:1}{a}. From this excited ${}^3E$ state, direct transitions into the phonon sideband (PSB) of the ground state ${}^3A_2$ are possible via spontaneous emission or stimulated emission induced by red probe-laser photons \cite{jeske2017stimulated}. Apart from direct optical transitions, an intersystem crossing (ISC) from ${}^3E$ into the excited singlet state ${}^1A_1$ can take place, followed by a fast weakly-emitting near-infrared (NIR) transition into the lower singlet state ${}^1E$ \cite{rogers2008iremission}. The ISC rate is higher when the ISC process originates from the $m_S=\pm1$ spin state of ${}^3E$. As a result, the transfer of population into the singlet states can be enhanced by a \SI{2.87}{GHz} microwave field resonant to the splitting of the spin states ($m_S=0\rightarrow m_S=\pm1$) in the ground state ${}^3A_2$ (see Fig.$\,$\fref{fig:1}{a}). This leads to an increased NIR absorption of the lower singlet state $^1E$. The ISC between the ${}^1E$ and ${}^3A_2$ states is more likely to transition into the $m_S=0$ spin state allowing to spin-polarize the NV center via optical excitation. These mechanisms are the foundation of absorption-based optically detected magnetic resonance (ODMR) \cite{acosta2010magnetometry}. Additionally, the combination of applying a resonant microwave field while optically pumping enables a spectroscopic investigation which selectively addresses NV$^-$ center effects when investigating only microwave-induced signal changes. Therefore, absorption and stimulated emission processes by other defects in the red and NIR spectral regime, which can occur upon optically pumping the diamond, are suppressed.

Our measurements exploit the enhancement of weak measurement signals by using a high-finesse optical cavity designed for a broadband red-to-NIR spectral range enabling detailed spectroscopic investigations. The experimental setup (see Fig.$\,$\fref{fig:1}{b}) consists of a \SI{532}{nm} pump laser and a tunable red-to-NIR probe laser (680-\SI{1065}{nm}) adjusted to a cavity input power of 0.1\,W throughout all experiments. Both laser beams are focused onto a high-pressure, high-temperature (HPHT) bulk diamond sample (\SI{1.97}{ppm} NV, 97$\,\%$ NV$^-$, \SI{24}{ppm} $\text{N}_{s}^0$). The diamond was pre-treated with a low-pressure, high-temperature (LPHT) treatment to decrease the absorption coefficient \cite{hahl2022thesis}. The sample is placed inside a cavity designed for the probe laser with a length of \SI{18.5}{mm}, formed by two identical mirrors with a radius of curvature of \SI{30}{mm} and a high reflectivity of >\,99.8\,\% over the investigated wavelength range. Through a microwave loop-antenna (diameter $\sim$\,\SI{3}{mm}), the microwave fields can be applied at the diamond.This antenna design requires high input powers, as the antenna cannot be impedance-matched, leading to high back-reflections at the ports.\\

In our setup, the cavity length is scanned with a piezo actuator attached to the rear cavity mirror \cite{hahl2022ltm, schall2025magnetometry}. The probe laser couples into the cavity only at specific cavity lengths which fulfill the standing-wave resonance criterion for the current probe wavelength. The transmitted probe-laser signal through the cavity is detected and displayed on an oscilloscope. We distinguish three experimental cases denoted by the indices
\begin{itemize}
    \item "Bare": probe on, pump off, microwave off 
    \item "Pump": probe on, pump on, microwave off 
    \item "MW": probe on, pump on, microwave on.
\end{itemize}

Cavity resonances for these three cases are shown in Fig.$\,$\fref{fig:2}{a}. The cavity signal is characterized by its amplitude $A$, which represents the power of the transmitted signal, its free spectral range (FSR), and its full width at half maximum (FWHM), which are indicated for the bare signal (gray) in Fig.\,\fref{fig:2}{a}.
Scanning the cavity length and using a stable narrow-linewidth Ti:sapph probe laser allows to determine the cavity finesse $\mathcal{F}=\text{FSR}/\text{FWHM}$ directly from the recorded cavity resonances \cite{hahl2022ltm}. The cavity finesse $\mathcal{F}$ is a measure of the inverse cavity losses and is therefore strongly probe wavelength dependent \cite{schall2025magnetometry}.\\ 

Throughout the entire spectral range of the probe laser, a high finesse $\mathcal{F}\sim500-1250$ of the cavity leads to an increased effective absorption length of the probe-laser light. This increase is quantified by the cavity-enhancement factor $\mathcal{G}=2\mathcal{F}/\pi$ \cite{orr2024cavityenhanc}.\\
We define the quasi-single-pass signal (QSP) changes $\alpha$ as
\begin{align}
    \alpha=\left(\frac{A_{\text{Bare}}-A_{\text{Pump}}}{A_{\text{Bare}}}\right)/ \:\mathcal{G}_{\text{Bare}}.\label{eq:Absorption}
\end{align}
These changes can either represent an amplification ($\alpha<0$) or an absorption ($\alpha>0$) with respect to the bare cavity signal. To investigate the pump-induced signal changes, the bare cavity signal is chosen as the reference point and is used to normalize the measurement signal changes with the corresponding cavity-enhancement factor $\mathcal{G}_{\text{Bare}}$. This normalization by the cavity-enhancement factor is essential to correct for different cavity-enhancements at different probe wavelengths due to the wavelength dependence of the finesse. It establishes the comparability of measurements over the entire probe laser range and yields values corresponding to the equivalent signal changes in a single-pass configuration.\\

\begin{figure*}[bt]
    \centering
    {\includegraphics[]{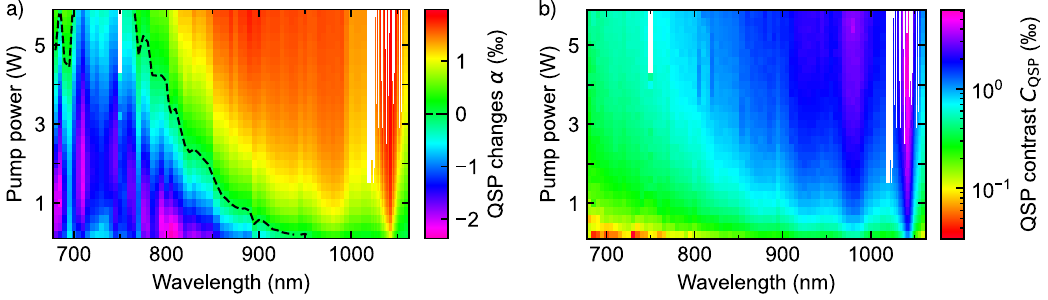}}
    \caption{\textbf{Quasi-single-pass pump-induced signal changes and collective ODMR contrast.} a) Pump-induced signal changes $\alpha$ over the entire available spectral and pump power range normalized to a quasi-single-pass (QSP) following Eq.\,(\ref{eq:Absorption}). The black dotted line separates amplification and absorption regimes which are mainly associated with NV$^-$ triplet stimulated emission and singlet absorption, respectively. The strongest absorption occurs around the singlet ZPL at \SI{1042}{nm}. Between 800-\SI{950}{nm} triplet emission and singlet absorption regimes overlap, leading to a competition of amplification and absorption effects. Which one of these effects dominates overall is dependent on the pump power. b) QSP collective ODMR contrast over the entire available spectral and pump power range using a microwave power of \SI{49}{dBm}. The ODMR contrast is an NV$^-$-selective quantity excluding contributions of other defects and was calculated using Eq.\,(\ref{eq:Contrast}). A similar NIR structure compared to a) is observed and assigned to the singlet transition and its phonon sideband. The highest ODMR contrasts are achieved in this singlet absorption regime, significantly exceeding the ODMR contrasts available through triplet stimulated emission. White tiles mark filtered areas with faulty experimental data.}
    \label{fig:3}
\end{figure*}

An exemplary measurement of the QSP signal changes $\alpha$ and the cavity parameters for varying pump powers is depicted in Fig.$\,$\fref{fig:2}{b}. There is a regime of amplification ($\alpha<0$) up to a pump power of $\sim\SI{0.6}{\watt}$. Above this threshold, the pump-laser-induced absorption exceeds the gain, leading to an overall absorption of the probe laser ($\alpha>0$). The gain (losses) can be attributed to stimulated emission (singlet absorption) of the NV centers \cite{hahl2022ltm, schall2025ltm}. However, these effects cannot fully explain the observed trends (see Sec.\,SI for details). Except for the used sign convention, the same trend portrayed in $\alpha$ is visible in the amplitude $A$ and the finesse $\mathcal{F}$ of the cavity. The correlation between the amplitude and the finesse is known in literature \cite{gagliardi2014book}. Consequently, during the normalization of $\alpha$, only the bare cavity signal can be used in order to avoid an unwanted cancellation effect of the pump-induced measurement signal changes. Furthermore, the finesse trends are entirely encoded in changes in the FWHM, because the FSR is constant for varying pump powers. We use this property of the FSR to filter out faulty measurements.

In analogy to the QSP signal changes $\alpha$, the QSP ODMR contrast $C_{\text{QSP}}$ can be obtained from the cavity-enhanced ODMR contrast $C_{\text{enh}}$ through the following normalization
\begin{align}
    C_{\text{QSP}}=\left(\frac{A_{\text{Pump}}-A_{\text{MW}}}{A_{\text{Pump}}}\right)/ \:\mathcal{G}_{\text{Pump}}=C_{\text{enh}} / \:\mathcal{G}_{\text{Pump}}.\label{eq:Contrast}
\end{align}
In this case, microwave-induced relative deviations from the cavity signal with an active pump laser are studied. Hence, the reference point is the cavity signal with an active pump laser (index "Pump"), which is why the normalization is performed with the corresponding cavity-enhancement factor $\mathcal{G}_{\text{Pump}}$. This enhancement factor is strongly dependent on the probe wavelength and the pump power (see Sec.\,SII\,A for details). \\
All measurements were performed without an external bias magnetic field. Consequently, the observed ODMR contrast is a collective contrast caused from driving the $m_S=0\rightarrow m_S=\pm1$ ground state transitions of all four NV orientations simultaneously \cite{barry2020review}.

\section{Results}
\subsection{Pump-laser-induced signal changes}
To further spectrally investigate the complex signal changes upon pumping the NV centers, the exemplary measurement series shown in Fig.$\,$\fref{fig:2}{b} is repeated for varying probe wavelengths. The quasi-single-pass (QSP) signal changes are quantified according to Eq.$\,$(\ref{eq:Absorption}) and are depicted in Fig.$\,$\fref{fig:3}{a}. A black dotted line marks the boundary between an amplification ($\alpha<0$) and an absorption ($\alpha>0$) of the signal with respect to the bare cavity signal before the pump-laser was switched on. \\
The recorded QSP pump-induced signal changes can be divided into two main regimes: (i) an amplification region ($\sim$\,680-\SI{900}{nm}), and (ii) an absorption region ($\sim$\,800-\SI{1065}{nm}). However, the transition wavelength between these two regions strongly depends on the pump power.
The amplification region can be attributed to NV$^-$ triplet stimulated emission and the absorption region to NV$^-$ singlet NIR absorption. Both transitions encompass broadband features due to their respective associated PSB structures that spectrally overlap \cite{kehayias2013spectroscopy, fraczek2017stim_em_both_NV}. The strongest QSP absorption appears at \SI{1042}{nm}, corresponding to the ZPL of the NV$^-$ singlet transition. The measurement was repeated with an additional applied resonant microwave field thereby changing the NV$^-$ dynamics and yielding the microwave-induced QSP signal changes with active pump laser (see Sec.\,SI\,A for details).\\ 
However, there are other observable pump-induced features in Fig.$\,$\fref{fig:3}{a} that cannot be explained by NV$^-$ alone. These features include the amplification peaks at around \SI{680}{nm} and \SI{800}{nm}, which do not coincide with the maximum of NV$^-$ stimulated emission \cite{fraczek2017stim_em_both_NV}, and an unexpected pump-induced absorption at high pump powers and low wavelengths (see Sec.$\,$SI\,B-C for more details).
These effects make it clear that the cavity signal contains potentially relevant contributions beyond those of the NV$^-$ center. To avoid undesired pump-induced effects in the singlet absorption regime, we turn to an NV$^-$ selective parameter in the next section to further investigate the singlet absorption.

\subsection{ODMR contrast}
\begin{figure*}[t]
    \centering
    {\includegraphics[]{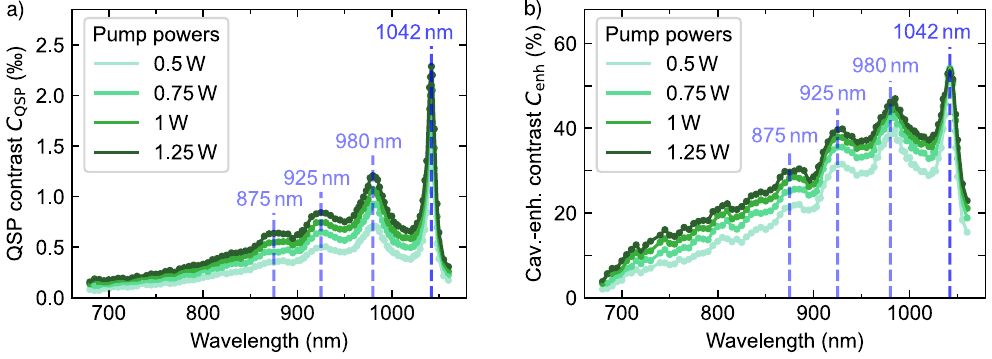}}
    \caption{\textbf{Quasi-single-pass and cavity-enhanced collective ODMR contrasts for various pump powers.} a) Quasi-single-pass (QSP) collective ODMR contrast $C_{\text{QSP}}$ at a microwave power of \SI{40}{dBm} revealing the room-temperature NV$^-$ singlet ${}^1E\rightarrow{}^1A_1$ absorption structure for higher wavelengths. The highest contrast appears at the ZPL at \SI{1042}{nm}. Broad peaks in the phonon sideband (\SI{980}{nm}, \SI{925}{nm}, \SI{875}{nm}) correspond to the energy spacing of an expected singlet phonon mode at $\sim$\SI{75}{meV} \cite{Jin2022SingletPhononTheo, kehayias2013spectroscopy}. b) Cavity-enhanced collective ODMR contrast $C_{enh}$ showing the data from a) without normalization by the cavity-enhancement factor (see Eq.\,(\ref{eq:Contrast})). The spectrum demonstrates very high possible room-temperature contrasts exceeding 50$\,\%$ caused by singlet absorption when increasing the effective absorption length with an optical cavity. The prominent and spectrally broad room-temperature phonon sideband enables high cavity-enhanced contrasts even at lower wavelengths than the ZPL.}
    \label{fig:4}
\end{figure*}

The ODMR contrast is the parameter of choice to investigate NV$^-$ effects only. It corresponds to the relative change of the cavity signal upon driving the NV$^-$ centers into their $m_S=\pm 1$ states, thereby creating a higher population in the NV singlet state (see Fig.\,\fref{fig:1}{a}). In Fig.$\,$\fref{fig:3}{b}, the QSP ODMR contrast $C_{\text{QSP}}$, as defined in Eq.$\,$(\ref{eq:Contrast}), is investigated spectrally and in dependence of pump power. The normalization by the cavity-enhancement factor provides QSP values for the ODMR contrast. To verify this method, we measured an actual single-pass contrast at \SI{1042}{\nano\metre} and a pump power of \SI{5}{\watt}. We find a very good agreement between the results in Fig.\,\fref{fig:3}{b} and the true single-pass measurement with comparable experimental parameters, which verifies the employed normalization method (see Fig.\,S3a).\\
The NIR structure appearing in the ODMR contrast in Fig.$\,$\fref{fig:3}{b} for wavelengths >\,\SI{800}{\nano\meter} is in good agreement with the NIR absorption structure in Fig.$\,$\fref{fig:3}{a}. Furthermore, the ODMR contrast in the regime of triplet stimulated emission is significantly lower than the ODMR contrast in the regime of singlet absorption. This indicates that in a continuous-wave pump-probe cavity setup, high contrasts can be more easily achieved with absorption-based magnetometry. The higher ODMR contrast in the singlet absorption is a direct result of a significantly longer lifetime and a larger population in the ${}^1E$ state than in the ${}^3E$ state \cite{acosta2010singlet} (see Fig.\,S5b). Additionally, multiple singlet absorption events can occur for the same NV$^-$ center before it returns to its ground state.\\
In Fig.\,\fref{fig:3}{b}, some peaks besides the ZPL of the singlet transition are observable. To resolve these features more clearly, the measurement was repeated for selected pump powers and with better statistics. The resulting spectrum of the QSP ODMR contrast $C_{\text{QSP}}$ is shown in Fig.$\,$\fref{fig:4}{a}.
At wavelengths exceeding the NV$^-$ triplet emission, the QSP ODMR contrast reveals the exact room-temperature singlet absorption structure. Several broad peaks are visible in Fig.$\,$\fref{fig:4}{a}, including the ZPL at \SI{1042}{nm} and further peaks around \SI{980}{nm}, \SI{925}{nm}, and \SI{875}{nm} corresponding to the singlet PSB structure. The energy difference of these peaks amounts to multiples of a 75\,meV phonon. This energy is in good agreement with a reported phonon mode believed to be a quasi-local mode of ${}^1A_1$ \cite{kehayias2013spectroscopy, Jin2022SingletPhononTheo}. The highest contrast is observed at \SI{1042}{nm}, which means that apart from technical limitations, this wavelength is best suited for absorption-based NV magnetometry. An additional strong improvement in ODMR contrast is observed for increasing pump powers, due to a better spin preparation in $m_S=0$. However, for higher pump powers, a saturation in the ODMR contrast occurs from the combination of saturating the triplet transition and ionizing the NV$^-$ centers (see Fig.\,S5).\\

The shown broadband absorption of the NV$^-$ singlet transition is particularly relevant for absorption-based two-media LTM, as it enables new gain-medium material systems to be used as active media. Although the absorption at \SI{1042}{\nano\metre} is strongest, utilizing a singlet phonon sideband for LTM offers several potential advantages. In particular, it allows the use of high-performance silicon detectors with high responsivity. Furthermore, the absorption varies less strongly with wavelength because the spectral features in the PSB are broader and less pronounced than the sharp absorption peak at \SI{1042}{\nano\metre}. Consequently, the translation of frequency noise to intensity noise in the combined laser system is weaker as mode hopping due to spectrally varying NV$^-$ absorption strengths is less likely to occur, thereby relaxing the constraints on frequency selection in LTM systems.\\
To demonstrate the potential of absorption- and cavity-based NV magnetometry, the cavity-enhanced ODMR contrast $C_{\text{enh}}$ is displayed in Fig.$\,$\fref{fig:4}{b}. This representation is achieved when omitting the normalization by the cavity-enhancement factor in Fig.$\,$\fref{fig:4}{a}. At the ZPL, the collective ODMR contrast in this spectrum exceeds 50$\,\%$. A zero-field cavity ODMR measurement (see Fig.$\,$S3b) and an investigation of the pump power dependence of the cavity-enhanced ODMR contrast (see Fig.$\,$S4) illustrate collective ODMR contrasts even surpassing 60$\,\%$ and reaching up to 80$\,\%$. We note here that further improvements in the cavity-enhanced ODMR contrast can be achieved by systematically optimizing the involved powers, increasing the effective absorption length or using a higher NV$^-$ concentration.\\

\subsection{Singlet absorption cross section}
\begin{figure*}[htb]
    \centering
    \includegraphics{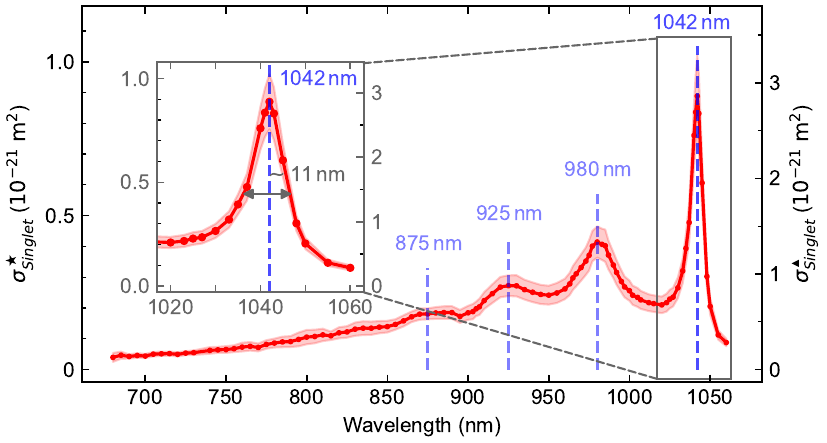}\\[1em]
    \caption{\textbf{Room-temperature absorption cross section (ACS) for the NV$^-$ singlet transition with its corresponding phonon sideband.} The depicted curve characterizes the ${}^1E\rightarrow{}^1A_1$ transition and corresponds to an average of multiple ACSs obtained from several combinations of pump and microwave powers, which we use to cross-validate our experimental method (see Fig.$\,$S10). The left and right y-axes indicate the corresponding singlet ACS values depending on which \SI{532}{nm} NV$^-$ ACS was used ("$\bigstar$"$\rightarrow$ \cite{abs532_0_95} and "$\blacktriangle$"$\rightarrow$ \cite{wee2007sigma532_0_31}). At the ZPL, located at \SI{1042}{nm} and displayed in the inset, an ACS of $\sigma^{\,\bigstar}_{1042}=(0.89\pm0.14)\cdot 10^{-21}\,\text{m}^2$ or $\sigma^{\,\blacktriangle}_{1042}=(2.9\pm0.5)\cdot 10^{-21}\,\text{m}^2$. and a FWHM of $\sim$\,\SI{11}{nm} are found.  A quasi-local \SI{75}{meV} phonon mode leads to a pronounced PSB with peaks at \SI{980}{nm}, \SI{925}{nm} and \SI{875}{nm}. The selected measurement method includes all \SI{2.87}{GHz}-microwave-induced cavity signal changes. Therefore, a microwave-induced reduction of triplet stimulated emission leads to an undesired contribution at lower wavelengths, resulting in an overestimation of the singlet ACS. Hence, at lower wavelengths, the measurement only corresponds to an upper bound of the singlet ACS.}
    \label{fig:5}
\end{figure*}
The ${}^1E\rightarrow{}^1A_1$ absorption cross section (ACS) $\sigma$ represents a concrete and comparable quantity for the singlet absorption strength. It is determined with a method which relies on the additional singlet absorption induced by the resonant microwave field, utilizing the equation
\begin{align}
    \sigma=\frac{V_{\text{Probe}}\cdot\mu}{\Delta N_{\text{Singlet}}}. \label{eq: abs cs}
\end{align}
In Eq.$\,$(\ref{eq: abs cs}), $V_{\text{Probe}}$ corresponds to the probed diamond volume, $\Delta N_{\text{Singlet}}$ represents the number of additional singlet absorbers induced by the microwave field, and $\mu$ is the corresponding absorption coefficient.\\
The absorption coefficient $\mu$ is determined with a Beer-Lambert approach, using measurements of the cavity amplitudes $A_{\text{Pump}}$ and $A_{\text{MW}}$ as well as the effective absorption length $L$:
\begin{align}
    A_{\text{MW}}=A_{\text{Pump}}\cdot e^{-\mu\cdot L},\\
    L=L_0\cdot\mathcal{G}_{\text{Pump}}.
\end{align}
Here, $L_0=322\,$µm corresponds to a single-pass through the diamond at Brewster's angle.\\
To determine the additional number of singlet absorbers induced by the microwave field $\Delta N_{\text{Singlet}}=N_{\text{MW}}-N_{\text{Pump}}$, a rate model including both NV charge states is set up (see Fig.$\,$S7). The steady-state populations $N_{\text{Pump}}$ and $N_{\text{MW}}$ are determined by solving the equations of motion resulting from the corresponding Lindblad master equation with and without resonant microwave field at the same pump excitation rates. The Rabi frequencies $\Omega$ are measured for identical experimental conditions with respect to the absorption coefficient measurements, using lock-in detected continuous-wave Rabi-measurements with the photoluminescence signal as detection signal (see Sec.\,SIII\,B).
A parameter which significantly affects the singlet ACS result is the pump excitation rate $\Lambda_{\text{NV}}$, which is related to the ACS of NV$^-$ at \SI{532}{nm}. There are two different established values for the \SI{532}{nm} NV$^-$ ACS in literature which we denote as $\sigma^{\,\bigstar}_{532}=(9.5\pm2.5)\cdot 10^{-21}\,\text{m}^2$ \cite{abs532_0_95} and $\sigma^{\,\blacktriangle}_{532}=(3.1\pm0.8)\cdot 10^{-21}\,\text{m}^2$ \cite{wee2007sigma532_0_31}. Singlet ACSs were determined for both of these NV$^-$ pump ACSs.
Further information on the ACS determination is presented in Sec.\,SIII.\\
To cross-validate our experimental method, the corresponding $\mu$ and $\Delta N_{\text{Singlet}}$ are determined for a set of pump and microwave powers by calculating several values of the ACS $\sigma$. All values are in good agreement with each other (see Fig.$\,$S9). Taking into account all the individual ACSs, each resulting from a distinct pair of pump- and microwave powers, the average and standard deviation is determined and presented in Fig.$\,$\ref{fig:5}. The ${}^1E\rightarrow{}^1A_1$ room-temperature ACS values at the absorption peaks are listed in Tab.\,\ref{tab:crosssections}.

\renewcommand{\arraystretch}{1.5}
\begin{table}[b]
\centering
\caption[Singlet absorption cross sections]{\textbf{Singlet ACSs at the ZPL and PSB absorption peaks.} Two separate evaluations are made using different literature values for the \SI{532}{nm} pump ACSs $\sigma^{\,\bigstar}_{532}$ \cite{abs532_0_95} and $\sigma^{\,\blacktriangle}_{532}$ \cite{wee2007sigma532_0_31}.}
\label{tab:crosssections}
\begin{tabular}{|l|l|}
\hline
ACS $\bigstar$ ($10^{-21}\,\text{m}^2$)& ACS $\blacktriangle$ ($10^{-21}\,\text{m}^2$) \\ \hline
 $\sigma^{\,\bigstar}_{532}= 9.5$& $\sigma^{\,\blacktriangle}_{532}=3.1$  \\ \hline
 $\sigma^{\,\bigstar}_{1042}=0.89\pm0.14$& $\sigma^{\,\blacktriangle}_{1042}=2.9\pm0.5$  \\ \hline
 $\sigma^{\,\bigstar}_{980}=0.41\pm0.06$& $\sigma^{\,\blacktriangle}_{980}=1.34\pm0.17$  \\ \hline
 $\sigma^{\,\bigstar}_{925}=0.27\pm0.04$& $\sigma^{\,\blacktriangle}_{925}=0.88\pm0.12$  \\ \hline
 $\sigma^{\,\bigstar}_{875}=0.181\pm0.018$& $\sigma^{\,\blacktriangle}_{875}=0.58\pm0.07$  \\ \hline
\end{tabular}
\end{table}

The FWHM of the absorption peak around the ZPL corresponds to about \SI{11}{nm}.\\
With our measurement method, undesired contributions of microwave-induced NV$^-$ triplet stimulated emission changes are present at lower wavelengths. These contributions enter the ACS calculation through the measured cavity signal amplitudes $A_{\text{MW}}$ and $A_{\text{Pump}}$ used to calculate the absorption coefficient $\mu$. They lead to an overestimation of the singlet ACS value at lower wavelengths where the ACS shown in Fig.\ref{fig:5} therefore has to be regarded as an upper bound.\\
Our measurements of the ZPL peak width and the ratio of the absorption strengths at the ZPL and the first PSB peak (\SI{980}{nm}) agree with reports from Kehayias et al. \cite{kehayias2013spectroscopy}. However, their reported room-temperature ACS is smaller, which they have determined to be $\sigma_{1042}^{\text{Keh}}=(0.40\pm0.09)\cdot 10^{-21}\,\text{m}^2$. We attribute the deviation to expected differences in the specifics of the underlying rate model calculations, which are not explained in their report, and a possible inclusion of undesired non-NV-related pump-induced effects in their experimental method. Another report by Dumeige et al. \cite{dumeige2013magnetometry} determined the ACS to be $\sigma_{1042}^{\text{Dum}}=(0.20\pm0.03)\cdot 10^{-21}\,\text{m}^2$. Their calculation method is based on numerically solving differential equations for the intensities used to describe an approximate ODMR contrast measurement at \SI{1042}{nm} reported in \cite{acosta2010magnetometry}. These differential equations include singlet populations from a rate model calculation which differs from our model as it does not include NV$^0$ and uses slightly different rates for several NV$^-$ transitions. The above described differences could result in deviations of the final ACS value.\\
We further investigated the influence of the individual rate model parameters on the absorption cross section result. A simultaneous \SI{10}{\%} relative error on all rate model parameters given in Tab.\,S1 except for $\Lambda_{\text{NV}}$ and $N_{\text{NV}}$ yields an uncertainty of only 12-15\,\% on the singlet ACS depending on the utilized pump- and microwave powers and pump excitation rate. In contrast to that, a \SI{10}{\%} relative error on the pump excitation rate $\Lambda_{\text{NV}}$ alone yields an uncertainty of already 21-29\,\% on the singlet ACS, which justifies the special focus put on this parameter during our evaluation. Another potential source of uncertainty lies within the exact determination of the number of NV centers $N_{\text{NV}}$, which has a linear influence on the absorption cross section (see Sec.\, SIII\,A and Eq.\,(\ref{eq: abs cs})).

\section{Discussion}
We have investigated the room-temperature NV$^-$ singlet ${}^1E\rightarrow{}^1A_1$ absorption including its broad phonon sideband and determined the associated ACSs using a novel method relying on microwave-induced cavity signal changes, thereby going beyond previous studies \cite{kehayias2013spectroscopy, younesi2022broadband}. We detected a broadband pump-induced amplification in the red and a broadband pump-induced absorption in the NIR spectral region. These features can be mainly explained by NV$^-$ triplet stimulated emission and singlet absorption processes, respectively.  We further used resonant microwave fields to determine the broadband ODMR contrast as an NV$^-$-selective quantity. Due to a lack of other spin-state-dependent NIR absorbing transitions within NV$^-$, the emerging NIR structure observed in both the pump-induced absorption and ODMR contrast can thus be assigned to NV$^-$ singlet absorption. Further evidence is provided by the observed spectral features including the singlet ZPL at \SI{1042}{nm} and multiple peaks coinciding with the energy spacing of an expected ${}^1A_1$ quasi-local phonon mode. The range of absorption magnetometry overlaps with the phonon sideband of emission magnetometry. While the NV$^-$ gain and NV$^-$ absorption induced by the \SI{532}{nm} pump laser work in opposite directions, the magnetic contrasts of both effects are in the same direction (lower signal on magnetic resonance) and thus add up. This contributes to the overall strong observable contrasts in the entire wavelength range from $\sim$\,680-1050\,nm when using cavity-enhanced magnetometry or LTM.\\
We were able to determine room-temperature ACS values for the singlet transition throughout all of its phonon sideband. They enable easy comparison of the absorption strength with other defect-photon interaction processes in diamond. The highest room-temperature ACS occurs at \SI{1042}{nm} with an FWHM of $\sim$\,\SI{11}{nm}. Depending on the employed NV$^-$ \SI{532}{nm} ACS, it corresponds to $\sigma^{\,\bigstar}_{1042}=(0.89\pm0.14)\cdot 10^{-21}\,\text{m}^2$ or $\sigma^{\,\blacktriangle}_{1042}=(2.9\pm0.5)\cdot 10^{-21}\,\text{m}^2$.\\
We have further shown that in our cavity setup, the NV$^-$ singlet absorption provides significantly better ODMR contrasts than the NV$^-$ triplet stimulated emission. Our results demonstrate the effectiveness of absorption-based NV magnetometry when enhancing the absorption length through an optical cavity, enabling collective ODMR contrasts in the range of 50-80\,\%. Although the highest ODMR contrasts are recorded around the singlet ZPL at \SI{1042}{nm}, high-contrast magnetometry at room temperature is possible throughout the entire PSB of the singlet transition. The probe wavelength thus becomes an additional degree of freedom to address technical limitations in absorption-based NV magnetometry. In particular, for absorption-based LTM, the phonon sideband offers several advantages including a potentially improved mode stability, greater flexibility in the emission wavelength of the gain medium and operation at a higher quantum efficiency when using silicon photodetectors.

\section*{Acknowledgments} 
T.P., F.S., J.J.S., L.L. and J.J. acknowledges the support from the German Federal Ministry of Research, Technology and Space, Bundesministerium für für Forschung, Technologie und Raumfahrt (BMFTR) under grant no. 13N16485. P.R. acknowledges support through Australian Research Council (ARC) Discovery Projects (DP220102518, DP250100125, DP260101975), an RMIT University Vice-Chancellor’s Senior Research Fellowship, and support from the US Air Force Office of Scientific Research (FA9550-25-1-0260). We thank F.\,A.\,Hahl, T.\,Luo, X.\,Vidal and H.\,Hapuarachchi for their contributions to earlier related work.

\section*{Data Availability}
The data that support the findings of this article are not publicly available. The data are available from the authors upon reasonable request.
\newpage
\onecolumngrid
\begin{center}
{\fontsize{14}{18}\selectfont\textbf{Supplementary materials for "Characterizing the nitrogen-vacancy center singlet transition and its phonon sideband for absorption-based room-temperature magnetometry"}}
\end{center}
\begin{center}
Tobias Probst\textsuperscript{1,*}, Florian Schall\textsuperscript{1,*}, Sebastian Heuft\textsuperscript{1}, Janina J. Schindler\textsuperscript{1},\\
Lukas Lindner\textsuperscript{1}, Sven Mädeffessl\textsuperscript{1}, Rüdiger Quay\textsuperscript{1}, Philipp Reineck\textsuperscript{2},
Alexander\\ 
M. Zaitsev\textsuperscript{3}, Takeshi Oshima\textsuperscript{4}, Matthias Weidemüller\textsuperscript{5}, and Jan Jeske\textsuperscript{1,$\dagger$}
\end{center}

\vspace{0.5em}

\begin{center}
\textit{
\textsuperscript{1}Fraunhofer Institute for Applied Solid State Physics IAF, Tullastr. 72, 79108 Freiburg, Germany\\[0.2em]
\textsuperscript{2}School of Science, RMIT University, Melbourne, VIC 3001, Australia\\[0.2em]
\textsuperscript{3}College of Staten Island (CUNY), Victory Blvd, Staten Island, 10312, New York, USA\\[0.2em]
\textsuperscript{4}National Institutes for Quantum Science and Technology (QST), 1233 Watanuki, Takasaki, Gunma 370-1292, Japan\\[0.2em]
\textsuperscript{5}Physikalisches Institut, Universität Heidelberg, Im Neuenheimer Feld 226, 69120 Heidelberg, Germany
}
\end{center}

\vspace{1em}

\begin{center}
This document contains sections SI to SIII, figures S1 to S10 and table S1.
\end{center}
\renewcommand{\thefigure}{S\arabic{figure}}
\setcounter{figure}{0} 
\renewcommand{\thetable}{S\arabic{table}}
\setcounter{table}{0}
\renewcommand{\thesection}{S\Roman{section}}
\setcounter{section}{0}
\section{Absorption and amplification mechanisms}

\subsection{Consequences of a resonant microwave field on the cavity signal}
By applying an additional resonant 2.87\,GHz microwave field to the experimental configuration depicted in Fig.$\,$\fref{fig:S1}{a} (identical to Fig.3a),  Fig.$\,$\fref{fig:S1}{b} is obtained, which shows the simultaneous display of the pump- and microwave-induced quasi-single-pass (QSP) cavity signal changes. Consequently, differences between Figs.\,\fref{fig:S1}{a} and \fref{fig:S1}{b} are directly caused by an increased NV$^-$ singlet population, a decreased triplet population, and a decreased effective triplet excited state lifetime due to an enhanced $m_S=\pm1$ population \cite{stuerner2021ms_lifetimes, Lin2025ms_lifteimes_nano}. The amplification at lower wavelengths is weaker in Fig.$\,$\fref{fig:S1}{b}, and the boundary between amplification and absorption is shifted towards lower wavelengths, while the absorption at higher wavelengths becomes stronger. These changes support the assignment of the absorption to the NV$^-$ singlet transition and the amplification to stimulated emission from the NV$^-$ triplet transition.\\

\subsection{Amplification peaks in absorption measurements}
\begin{figure*}[tb]
    \centering
    {\includegraphics[]{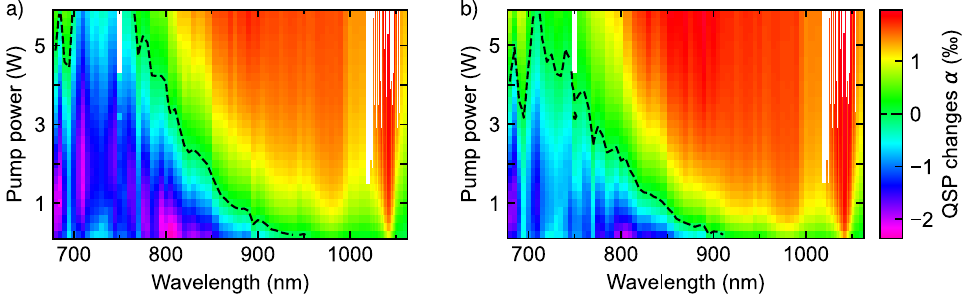}}
    \caption{\textbf{Measurements of quasi-single-pass signal changes $\alpha$.} a) Pump-induced quasi-single-pass (QSP) signal changes presented as Fig.$\,$3a in the main text. Negative signal changes correspond to amplification, positive signal changes correspond to absorption. A black dotted line marks the boundary between amplification and absorption. b) Microwave-induced QSP signal changes while the pump laser is turned on. To determine these QSP signal changes, $A_{\text{Pump}}$ in equation (1) from the main text is replaced with the amplitude $A_{\text{MW}}$. The difference to a) consists of the additional application of a \SI{2.87}{GHz} microwave field. Therefore, differences between a) and b) are caused by an increased NV$^-$ ${}^1E$ and decreased ${}^3E$ population. As a consequence, the low-wavelength amplification regime declines and the high-wavelength absorption regime increases. The black dotted boundary shifts to lower wavelengths. Furthermore, non-NV related effects become more evident through the different microwave-affected behaviors of the stimulated emission peaks and a pump-induced absorption at low wavelengths and high pump powers.}\label{fig:S1}
\end{figure*}

The strongest amplifications in Fig.$\,$\fref{fig:S1}{a} are observed at \SI{680}{nm}, \SI{710}{nm}, and \SI{800}{nm}. The wavelength of \SI{710}{nm} coincides with the peak of NV$^-$ stimulated emission \cite{fraczek2017stim_em_both_NV}. The other two wavelengths are likely attributable to stimulated emissions by defects other than NV$^-$. Upon resonant microwave excitation (see Fig.$\,$\fref{fig:S1}{b}), the dominant NV$^-$ stimulated emission decreases. This leads to an overall decrease in the amplification background at lower wavelengths and the disappearance of the peak at \SI{710}{nm}. However, the peaks at \SI{680}{nm} and \SI{800}{nm} are still visible in Fig.\,\fref{fig:S1}{b}. We assign the peak at \SI{680}{nm} to NV$^0$ stimulated emission in accordance with its reported stimulated emission peak \cite{fraczek2017stim_em_both_NV}. The peak around \SI{800}{nm} could be assigned to stimulated emission from the neutral monovacancy V$^0$, which has a reported broad stimulated emission peak around \SI{800}{nm} \cite{subedi2021V0stimem}. While the concentration of monovacancies is unknown and their absorption cross section (ACS) at \SI{532}{nm} is rather low, their peak stimulated emission cross section is significantly higher than the NV center's stimulated emission cross sections \cite{subedi2021V0stimem, fraczek2017stim_em_both_NV}. At higher pump powers, ionization of the NV centers (or other defects) increases the availability of electrons that can be captured to convert V$^0\rightarrow$V$^-$, which is active in the ultra-violet regime \cite{davies1992vacancy}, thereby leading to the demise of V$^0$ stimulated emission and a vanishing of the 800\,nm amplification peak. Furthermore, the role of monovacancies might have been underestimated previously due to their spontaneous emission being suppressed by non-radiative transitions at room temperatures leading to a reduction in PL \cite{davies1987radiativeGR1}.\\

\begin{center}
\rule{3cm}{0.4pt}
\end{center}

\vspace{1em}

\noindent
\textsuperscript{*}These authors contributed equally to this work.\\
\textsuperscript{$\dagger$} \texttt{jan.jeske@iaf.fraunhofer.de}

\newpage

\subsection{Pump-induced absorption at lower wavelengths}
At high pump powers, two-photon ionization of the NV center occurs \cite{aslam2013twophotonionization, chen2013twophotonionization532nm}, leading to a saturation or decrease of the stimulated emission from the NV$^-$ center. Simultaneously, free electrons could lead to a charge-state conversion of the monovacancy (V$^0\rightarrow$V$^-$). As a result, the dominant stimulated emission processes could decline, resulting in a weaker amplification at the respective wavelength regions. Nonetheless, the observed absorption at high pump powers and lower wavelengths (see Fig.\,\ref{fig:S1}) cannot be explained by these effects or other conventional means. We interpret this observation as another indication of non-NV-related, pump-laser-induced absorption \cite{hahl2022ltm}. A possible origin of this pump-induced absorption lies within the two-photon ionization of the NV$^-$ center with \SI{532}{nm} photons. This ionization provides an electron that can be captured by another defect, changing its charge state. This other defect could then become absorbing at the probe wavelength (e.\,g., $\text{N}_2{\text{V}}^0+e^-\rightarrow\text{N}_2{\text{V}}^-$). However, it is also possible that the two-photon charge-state conversion of the NV center is assisted by probe-laser photons \cite{hacquebard2018charge}. These photons could provide enough energy for the second step, i.\,e., the transitions from ${}^3E$ into the conduction band ($\text{N}{\text{V}}^-\rightarrow\text{N}{\text{V}}^0+e^-$) or from the valence band to the $a_1$ orbital of ${{}^2A}$ ($\text{N}{\text{V}}^0+e^-\rightarrow\text{N}{\text{V}}^-$) \cite{wood2024wavelength}. All these effects involve absorption of probe-laser photons and therefore change the cavity signal. 

\section{ODMR contrast}
\subsection{Cavity-enhancement factor}
\begin{figure*}[tb]
    \centering
    {\includegraphics[]{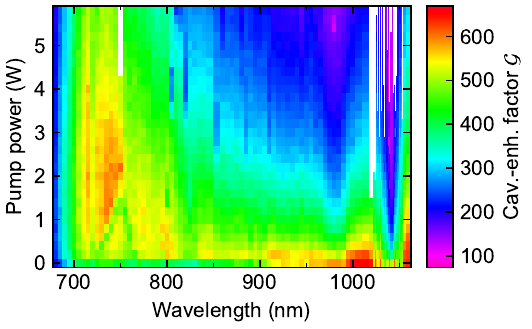}}
    \caption{\textbf{Quantification of the cavity-enhancement effect.} Cavity-enhancement factor $\mathcal{G}$ for varying probe wavelengths and pump powers showing a strong variation throughout the investigated parameter regimes. The data corresponds to the same measurement series depicted in Fig.\,3 in the main text and was used to obtain the quasi-single-pass contrasts depicted in Fig.\,3b in the main text.}\label{fig:cef}
\end{figure*}

The cavity-enhancement factor $\mathcal{G}$ for each configuration of the measurement series depicted in Fig.\,3 in the main text is shown in Fig.\,\ref{fig:cef}. A strong variation of $\mathcal{G}$, ranging from approximately 100 to 650, is observed across the investigated spectral and pump power regime. In addition, the same relative trends as for the QSP signal changes $\alpha$ are evident (see Fig.\,3a in the main text), as expected from the correlation between the cavity signal amplitude and the cavity finesse (see Fig.\,2b in the main text). Owing to the pronounced dependence on both pump power and wavelength, the ODMR contrast $C_{\text{enh}}$ must be normalized by the corresponding cavity-enhancement factor $\mathcal{G}_{\mathrm{Pump}}$ to remove the influence of varying cavity enhancement, as described by Eq.\,(2) in the main text. This normalization enables us to isolate the physical origin of the observed signal changes and represents a significant advance over our previous work in \cite{schall2025magnetometry}.

\begin{figure*}[b]
    \centering
    {\includegraphics[]{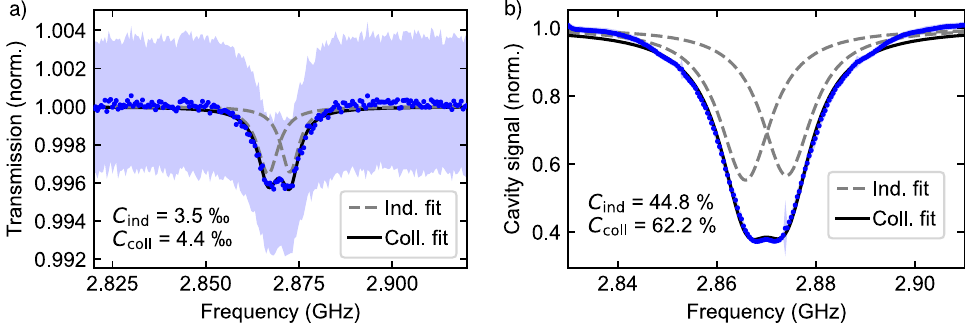}}
    \caption{\textbf{Absorption-based ODMR curves.} a) True single-pass measurement recording transmission changes at \SI{1042.5}{nm} with \SI{5}{W} pump power and \SI{43}{dBm} microwave power. The measured collective ODMR contrast is in good agreement with the expected values from the normalized cavity measurements depicted in Fig.$\,$3b in the main text. b) Representative cavity-enhanced ODMR measurement at \SI{1042}{nm} with a pump power of \SI{1.5}{W} and a microwave power of \SI{49}{dBm}, displaying high individual and collective ODMR contrasts. In both panels, the shaded region marks the standard deviation of the measurement.}\label{fig:S2}
\end{figure*}
\subsection{Justification of single-pass normalization}
To verify the normalization used to cancel out the cavity-enhancement effect, the experimental setup depicted in Fig.\,1a in the main text was modified by removing the two cavity mirrors. With this true single-pass configuration, an ODMR measurement is performed using the transmission of the probe laser as the readout signal (see Fig.$\,$\fref{fig:S2}{a}). For this measurement, the same diamond sample as in the main text is used and the probe-laser wavelength is fixed to \SI{1042.5}{nm}. The spot sizes of the lasers are comparable to those in the cavity experiments presented in the main text. The only experimental differences arise from a slightly decreased microwave power to \SI{43}{dBm} (previously \SI{49}{dBm}) and an antenna positioning that is not exactly reproducible. However, the achieved collective contrast $C_{\text{coll}}=4.4\,$‰ is in good agreement with the value of 5.6\,‰ obtained from Fig.$\,$3b in the main text. The slightly larger contrast is expected to result from the stronger microwave field. Consequently, the good agreement of both values justifies the applied QSP normalization (see Eq.\,(2) in the main text), which is used throughout the main text to account for differing cavity-enhancement effects.\\

\subsection{Cavity-enhanced ODMR contrast}
\begin{figure*}[tb]
    \centering
    \includegraphics{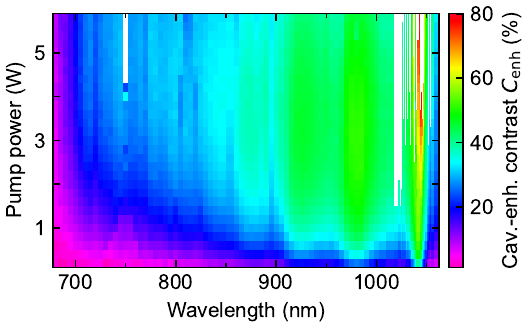}\\[1em]
    \caption{\textbf{Cavity-enhanced collective ODMR contrast.} The measurement corresponds to the same data shown in Fig.$\,$3b in the main text when omitting the cavity-enhancement normalization. It is recorded without external magnetic field over the entire available probe wavelength and pump power range using a microwave power of \SI{49}{dBm}. At the singlet ZPL at \SI{1042}{\nano\metre}, contrasts between $60-80\,\%$ can be achieved.}
    \label{fig:S3}
\end{figure*}

By using the cavity amplitude as a measurement signal, it is possible to record cavity-enhanced ODMR measurements. Fig.$\,$\fref{fig:S2}{b} shows such a measurement, yielding a collective contrast $C_{\text{coll}}=62.2 \,\%$. This demonstrates the potential of cavity- and absorption-based NV magnetometry due to their high achievable ODMR contrasts. We expect that the contrast can be further increased through improved cavity adjustment and a systematic optimization of all involved parameters. Indications of this are provided by further investigating the pump-power dependence of the collective ODMR contrast. If the normalization by the cavity-enhancement factor is omitted, the QSP collective ODMR contrast (see Fig.$\,$3b in the main text) can be reevaluated as the cavity-enhanced collective ODMR contrast, as depicted in Fig.$\,$\ref{fig:S3}. In certain areas of the investigated parameter space around \SI{1042}{\nano\metre}, collective ODMR contrasts reaching $80\,\%$ are found.

\subsection{Saturation effect with increasing pump power}
\begin{figure*}[tb]
    \centering
    {\includegraphics[]{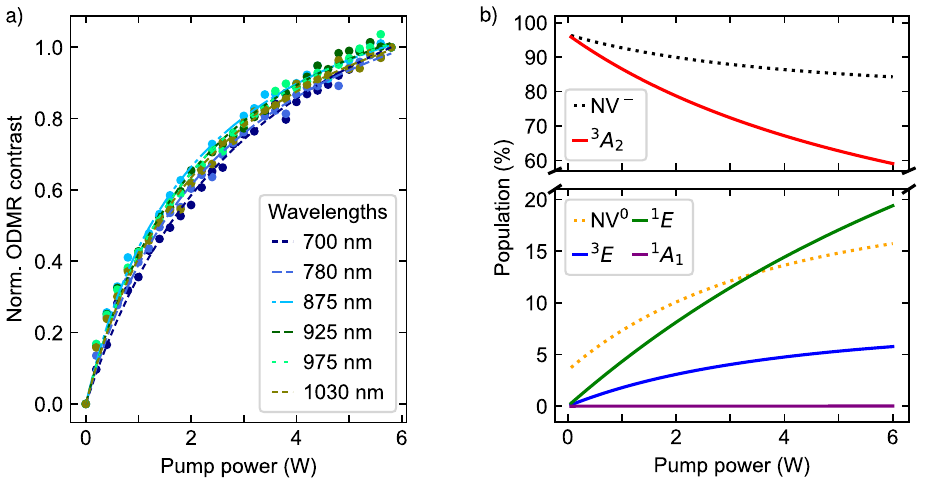}}
    \caption{\textbf{Saturation of collective ODMR contrast and modeled populations.} a) ODMR contrast normalized to its maximum for varying wavelengths over the entire probe wavelength range. A similar saturation behavior of the ODMR contrast with increasing pump power is observed for all investigated wavelengths. b) Steady-state populations for the total NV$^0$ and NV$^-$ populations and their relevant electronic states. With increasing pump power, ionization shifts the charge-state ratio in favor of NV$^0$. The individual NV$^-$ states demonstrate a saturation behavior in both the ${}^1E$ and ${}^3E$ populations, relevant for the ODMR contrasts obtained from triplet stimulated emission and singlet absorption. For the \SI{532}{nm} ACS, $\sigma^{\,\bigstar}_{532}$ \cite{abs532_0_95} was used.}\label{fig:S4}
\end{figure*}

The ODMR contrast measurements depicted in Fig.$\,$3b in the main text contain a saturation behavior with increasing pump power. This saturation appears to be similar for all probe wavelengths, with no clear distinction between the triplet stimulated emission spectral regime and the singlet absorption spectral regime, as visualized in Fig.$\,$\fref{fig:S4}{a}. The similarity of this saturation behavior in both the triplet stimulated emission and singlet absorption spectral regimes suggests a common origin. We attribute the observed saturation to a combination of two effects that directly impact the ${}^3E$ population and thereby also indirectly influence the ${}^1E$ population: (i) a weak onset of excitation saturation in the ${}^3A_2 \rightarrow {}^3E$ transition, and (ii) predominant ionization from the ${}^3E$ excited triplet state, which alters the NV charge-state ratio in favor of NV$^0$. Rate model calculations, including both NV charge states, confirm this ionization effect and a saturation effect in the relevant steady-state populations (see Fig.$\,$\fref{fig:S4}{b}).
Details on the calculation and the underlying rate model are given in Sec.\,\ref{sec:Ratemodel} .

\newpage
\section{Singlet absorption cross section}\label{sec:Abs_CS}
\subsection{Determination method and rate model}\label{sec:Ratemodel}
The following section describes the methods used to determine the ACS of the lower singlet state. The calculation is based on Eq.$\,$(3) in the main text, and its individual constituents are explained step by step below.\\

The probe volume $V_{\text{Probe}}$ is determined by assuming a cylindrical excitation volume with a length corresponding to the pass length through the diamond at Brewster's angle (\SI{322}{\micro\metre}) and a radius given by the cavity waist of \SI{67.5}{\micro\metre} at the diamond position.

\begin{figure*}[tb]
    \centering
    \includegraphics[]{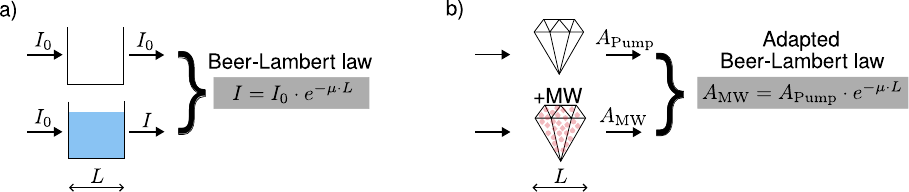}
    \caption{\textbf{Illustration of the experimental concept using a Beer-Lambert law.} a) The lower part illustrates the typical Beer-Lambert experiment, where the initial intensity $I_0$ decreases exponentially through an absorbing medium with absorption coefficient $\mu$ and length $L$ yielding the reduced intensity $I$. By considering an insertable absorbing medium, the intensities can be measured at a fixed detector position behind the absorbing medium with (lower part) and without (upper part) an inserted medium. b) Adapted Beer-Lambert law for our experimental situation. By applying a resonant microwave field (+MW), additional singlet absorbers (red dots) are activated in the diamond. We measure the cavity amplitudes before ($A_{\text{Pump}}$) and after ($A_{\text{MW}}$) the activation of these additional absorbers. An adapted Beer-Lambert law is formulated using the absorption coefficient $\mu$, assigned to these additional absorbers, and the effective absorption length, taking into account the cavity enhancement.}\label{fig:S5}
\end{figure*}

The absorption coefficient $\mu$ of the microwave-induced absorption is measured by determining the changes in the cavity amplitude $A$ upon applying a resonant microwave field at \SI{2.87}{\giga\hertz}. We apply the Beer-Lambert absorption law as illustrated in Fig.\,\ref{fig:S5}. This exponential law is typically used to determine the absorption losses of a laser with intensity $I_0$ passing through a medium of length $L$ and absorption coefficient $\mu$ \cite{martinez2021beerlambert}. For our purposes, we consider a modification of the typical experimental situation, using an insertable absorbing medium and a fixed detector position behind this medium. The intensities $I_0$ and $I$ can then be measured at the same detector position, with and without the inserted medium, respectively. In the actual experiment, we use the same principle. By applying a resonant microwave field, additional singlet absorbers are activated within the diamond, indicated by red dots in Fig.$\,$\fref{fig:S5}{b}, which is the equivalent of inserting the absorbing medium into the setup. We apply the Beer-Lambert law with the amplitudes $A_\text{Pump}$ and $A_{\text{MW}}$ representing the initial intensity and the intensity after passing through the diamond (see Eq.$\,$(4) in the main text). The absorption coefficient $\mu$ is then assigned to the absorption created by these additional singlet absorbers, i.\,e., the difference in the ${}^1E$ population $\Delta N_{\text{Singlet}}$ induced by a resonant \SI{2.87}{GHz} microwave field (see Eq.\,(\ref{eq:deltaNsinglet})). To account for the enhancement effect of the cavity, the length $L$ represents an effective absorption length and needs to be adapted following Eq.$\,$(5) in the main text using the cavity enhancement factor $\mathcal{G}_{\text{Pump}}=2F_{\text{Pump}}/\pi$. Thus, the absorption coefficient $\mu$ describing the microwave-induced additional absorption is given by 

\begin{equation}
    \mu=\frac{\ln\left(A_{\text{MW}}/A_{\text{Pump}}\right)}{L_0\,\mathcal{G}_{\text{Pump}}}
    \label{eq:mu}
\end{equation}

and can be determined by the cavity signal measurement procedure described in the main text. 

\begin{figure*}[tb]
    \centering
    {\includegraphics[]{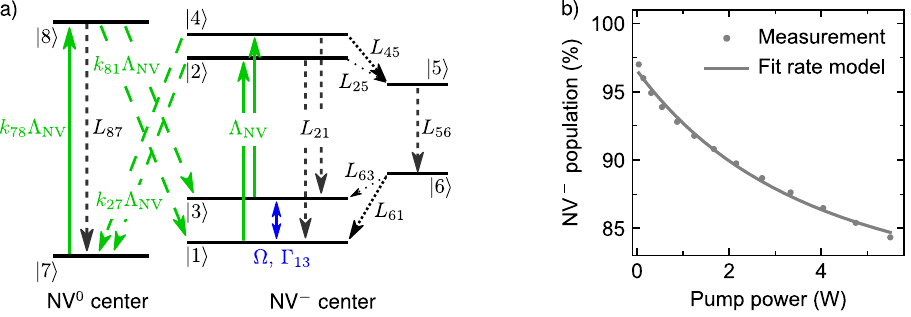}}
    \caption{\textbf{Rate model of the NV center and fitted ionization measurement.} a) NV center rate model including both NV charge states, enabling to model charge-state cycling through ionization and recombination dynamics. Coherent microwave driving is implemented via the Rabi frequency $\Omega$. The rates are summarized in Tab.$\,$\ref{tab:rates}. b) Ionization measurement from the same diamond used in this work adapted from \cite{hahl2022ltm}. The analytical total NV$^-$ population obtained through the solved Lindblad master equation, which is based on the rate model from a), is fitted to the measurement to determine the scaling factors $k_{27}$ (ionization), $k_{81}$ (recombination) and $k_{78}$ (NV$^0$ excitation).}\label{fig:S6}
\end{figure*}

Lastly, the number of additional NV centers populating the lower singlet state $\Delta N_{\text{Singlet}}$ upon applying a resonant microwave field is determined by solving the Lindblad master equation associated to this system. This method is based on previous work \cite{jeske2015ltm, rottstaedt2025ltm, schall2025ltm}, but to account for ionization processes and charge-state cycling, the existing model is expanded to also include the neutral charge state NV$^0$ of the NV center. With the help of this model, we find the analytical steady-state solution for the populations of all NV center electronic states, taking into account the optical excitation, ionization processes, and the coherent microwave drive. A schematic clarifying the following nomenclature and the involved electronic states and transitions is depicted in Fig.\,\fref{fig:S6}{a}. The equations of motion for the density matrix elements are given by 

\begin{gather}
    \dot{\rho}_{13}=(\Gamma_{13}-\Lambda_{\text{NV}})\rho_{13} + i\Omega(\rho_{11} -\rho_{33}),\\
    \dot{\rho}_{31}=-(\Gamma_{13}+\Lambda_{\text{NV}})\rho_{31} + i\Omega(\rho_{33} -\rho_{11}),\\
    \dot{\rho}_{11}= i\Omega(\rho_{13} -\rho_{31})-\Lambda_{\text{NV}}\rho_{11}+L_{21}\rho_{22}+L_{61}\rho_{66}+k_{81}\Lambda_{\text{NV}}\rho_{88},\\
    \dot{\rho}_{22}=\Lambda_{\text{NV}}\rho_{11} -(L_{21}+L_{25}+k_{27}\Lambda_{\text{NV}})\rho_{22},\\
    \dot{\rho}_{33}=i\Omega(\rho_{31}-\rho_{13})-\Lambda_{\text{NV}}\rho_{33}+L_{43}\rho_{44}+L_{63}\rho_{66}+k_{81}\Lambda_{\text{NV}}\rho_{88},\\
    \dot{\rho}_{44}=\Lambda_{\text{NV}}\rho_{33}-(L_{43}+L_{45}+k_{27}\Lambda_{\text{NV}})\rho_{44},\\
    \dot{\rho}_{55}=L_{25}\rho_{22}+L_{45}\rho_{44}-L_{56}\rho_{55},\\
    \dot{\rho}_{66}=L_{56}\rho_{55}- (L_{61}+L_{63})\rho_{66},\\
    \dot{\rho}_{77}=k_{27}\Lambda_{\text{NV}}\rho_{22}+k_{27}\Lambda_{\text{NV}}\rho_{44}+L_{87}\rho_{88},\\
    \dot{\rho}_{88}=k_{78}\Lambda_{\text{NV}}\rho_{77}-(L_{87}+2k_{81}\Lambda_{\text{NV}})\rho_{88}.
\end{gather}

The on-diagonal entries ($\rho_{11}$, ..., $\rho_{88}$) of the density matrix $\mathbf{\rho}$ correspond to the steady-state populations of the states $\ket1$ to $\ket8$ (see Fig.\,\fref{fig:S6}{a}). The terms $\rho_{13}$ and $\rho_{31}$ express coherences between the states $\ket{1}$ and $\ket{3}$. The rates $L_{ij}$ describe the decay rate from state $\ket{i}$ to $\ket{j}$. The coherent microwave drive is described by the Rabi frequency $\Omega$ and the decoherence rate is given by $\Gamma_{13}=1/T_2^*$. The optical excitation of the NV$^-$ centers is given by the excitation rate $\Lambda_{\text{NV}}$,

\begin{equation}
    \Lambda_{\text{NV}} = \sigma_{532} \cdot \frac{\text{$P$}_{\text{Pump}}}{\text{$E$}^{\text{ph}}_{532}\cdot \pi \text{$R$}_{\text{beam}}^2},
    \label{eq: LambdaNV}
\end{equation}

which is defined by the \SI{532}{nm} ACS of the NV$^-$ center $\sigma_{532}$, the photon energy $\text{$E$}^{\text{ph}}_{532}$ at \SI{532}{nm} and the pump beam radius $\text{$R$}_{\text{beam}}=$ 67.5$\,$µm on the diamond. There are two different established values for the \SI{532}{nm} NV$^-$ ACS in literature which we denote as $\sigma^{\,\bigstar}_{532}=(9.5\pm2.5)\cdot 10^{-21}\,\text{m}^2$ \cite{abs532_0_95} and $\sigma^{\,\blacktriangle}_{532}=(3.1\pm0.8)\cdot 10^{-21}\,\text{m}^2$ \cite{wee2007sigma532_0_31}. The influence of the probe laser is neglected due to the comparatively short lifetime (high decay rate $L_{56}$) of the upper singlet state and its therefore vanishing influence on the NV state populations for the probe powers used throughout this work \cite{acosta2010singlet, ulbricht2018NVIRlifetime}. The optical excitation of NV$^0$ as well as the ionization and recombination processes are described as multiples of the NV$^-$ excitation rate $\Lambda_{\text{NV}}$. The corresponding multipliers $k_{78}$ (NV$^0$ excitation), $k_{27}=k_{47}$ (ionization process), and $k_{81}=k_{83}$ (recombination process) are determined by fitting the analytical NV$^-$ population expression obtained from the rate model to an ionization measurement performed in previous work \cite{hahl2022ltm} with the same diamond sample. The results are depicted in Fig.\,\fref{fig:S6}{b}.\\

The analytical expressions for the steady-state population solutions are obtained by solving $\mathbf{\dot{\rho}}=0$ using Mathematica. Finally, the number $\Delta N_{\text{Singlet}}$ of additional NV centers populating the lower singlet state upon applying a resonant microwave field is determined via

\begin{equation}
    \Delta N_{\text{Singlet}}=\left(\rho_{66}(\Lambda_{\text{NV}}, \Omega)-\rho_{66}(\Lambda_{\text{NV}}, \Omega=0)\right)\cdot N_{\text{NV}}.
    \label{eq:deltaNsinglet}
\end{equation}

Here, $N_{\text{NV}}$ is the number of NV centers in the pumped volume $V_{\text{Probe}}$, which can be determined via
\begin{align}
    N_{\text{NV}}= \frac{[\text{NV}]_{\text{ppm}}}{10^6}\cdot n_{\text{C}}\cdot V_{\text{Probe}}.
\end{align}
We use $[\text{NV}]_{\text{ppm}}\approx$ \SI{1.97}{ppm} for the NV concentration and $n_C=8/a^3$ for the number density of carbon atoms in diamond, where $a$ corresponds to the diamond lattice constant \cite{Shikata2018diamondlatticeconstant}.
The Rabi frequencies $\Omega$ are determined via Rabi oscillations, as described in the next subsection. A complete list of all parameters used for the modeling in this work is given in Tab.\, \ref{tab:rates}. 

\renewcommand{\arraystretch}{1.2}
\begin{table}[tb]
\centering
\caption[Rates for rate model calculations]{\textbf{Rates and parameters used in the rate model calculations to determine the populations of individual NV-center states.}}
\label{tab:rates}
\begin{tabular}{|l|l||l|}
\hline
Parameter & Value & Reference  \\ \hline
 $L_{21}$& \SI{66.16}{MHz} & \cite{gupta2016timereolved}\\ \hline
 $L_{43}$& \SI{66.16}{MHz}  & \cite{gupta2016timereolved}\\ \hline
 $L_{25}$& \SI{11.1}{MHz} & \cite{gupta2016timereolved} \\ \hline
 $L_{45}$& \SI{91.8}{MHz}  & \cite{gupta2016timereolved}\\ \hline
 $L_{56}$& \SI{10}{GHz} & \cite{ulbricht2018NVIRlifetime} \\ \hline
 $L_{61}$& \SI{4.87}{MHz}  & \cite{gupta2016timereolved}\\ \hline
 $L_{63}$& \SI{2.04}{MHz}  & \cite{gupta2016timereolved}\\ \hline
 $L_{87}$& \SI{47.62}{MHz}  & \cite{storteboom2015NV0lifetime}\\ \hline
 $k^{\,\bigstar}_{27}$& 1.87& Fit Fig.\,\fref{fig:S6}{b} \\ \hline
 $k^{\,\bigstar}_{78}$& 0.78  & Fit Fig.\,\fref{fig:S6}{b}\\ \hline
 $k^{\,\bigstar}_{81}$& 19.74 & Fit Fig.\,\fref{fig:S6}{b}\\ \hline
 $k^{\,\blacktriangle}_{27}$& 15.90& analogous Fit\\ \hline
 $k^{\,\blacktriangle}_{78}$& 1.91  & analogous Fit\\ \hline
 $k^{\,\blacktriangle}_{81}$& 70.46 & analogous Fit \\ \hline
 $\Gamma_{13}$& \SI{5}{MHz}  & measured\\ \hline
$\Lambda^{\,\bigstar}_{\text{NV}}/\text{$P$}_{\text{Pump}}$& \SI{1.777}{\text{MHz}/\text{W}}  & Eq.\,(\ref{eq: LambdaNV}), \cite{abs532_0_95}\\ \hline
$\Lambda^{\,\blacktriangle}_{\text{NV}}/\text{$P$}_{\text{Pump}}$& \SI{0.580}{\text{MHz}/\text{W}}  & Eq.\,(\ref{eq: LambdaNV}), \cite{wee2007sigma532_0_31}\\ \hline
 $N_{\text{NV}} $& 1.602$\times\,10^{12}$  & measured\\ \hline
\end{tabular}
\end{table}

\newpage
\subsection{Continuous-wave Rabi measurements}
\begin{figure*}[b]
    \centering
    \includegraphics{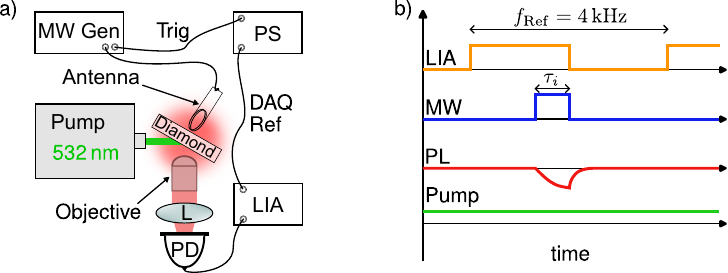}\\[1em]
    \caption{\textbf{Experimental procedure for continuous-wave Rabi measurements.} a) Experimental setup for photoluminescence (PL)-based Rabi oscillations when continuously pumping the NV centers. The PL signal is detected via an objective and focused on a photodetector (PD). The signal is analyzed via a lock-in amplifier (LIA), which gets a reference signal (Ref) and trigger signals for the data acquisition (DAQ) by a pulse streamer (PS). The PS additionally triggers (Trig) the microwave generator (MW Gen). b) Pulse sequences used for the cw Rabi protocol. The same sequence is repeated for varying pulse lengths $\tau_i$ resulting in Rabi oscillations after demodulation with the reference frequency $f_{\text{Ref}}$=\SI{4}{\kilo\hertz}.}
    \label{fig:S7}
\end{figure*}

\begin{figure*}[tb]
    \centering
    \includegraphics[]{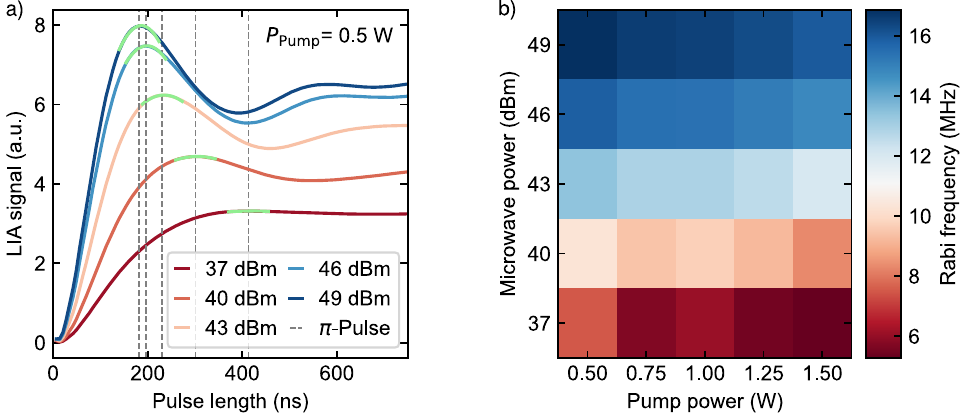}
    \caption{\textbf{Continuous-wave Rabi measurements without external magnetic field.} a) Representative Rabi measurement for a pump power of \SI{0.5}{W} showing the demodulated LIA signal as a function of the microwave pulse length for several microwave powers. The $\pi$-pulses are extracted from parabolic fits around the maxima of the first oscillation peaks indicated in green. b) Summary of the obtained Rabi frequencies for all measured pairs of pump and microwave powers. Each pair is used for an independent determination of the ACS. All measurements were recorded with a \SI{2.87}{GHz} microwave drive.}
    \label{fig:S8}
\end{figure*}

Rabi oscillations are performed to determine the Rabi frequencies necessary to estimate the population changes in the lower NV singlet state upon applying a resonant microwave (MW) field. The previously employed measurement procedure of scanning the cavity length can not be used for Rabi measurements because it generates a pulsed instead of a continuous-wave measurement signal. Therefore, we use the photoluminescence (PL) signal of the NV center as the measurement signal as illustrated in Fig.\,\fref{fig:S7}{a}. Unlike in the common NV Rabi protocols, where the pump laser is pulsed, we continuously keep it switched on as already proposed in \cite{zhang2021cwrabi}. The pulse sequence we utilize is shown in Fig.\,\fref{fig:S7}{b}. The photoluminescence is collected via an objective and focused onto the photodetector (PD) with a lens (L). A lock-in amplifier (LIA) is used to reference the measurement signal (with an applied MW pulse of length $\tau_i$) against the dark signal (without an applied MW pulse). A pulse streamer (PS) is employed to provide the reference signal (Ref, $f_{\text{Ref}}=\SI{4}{kHz}$) and the signals for the data acquisition (DAQ) of the LIA. Additionally, the PS triggers (Trig) the MW pulses at the MW generator (MW Gen). The sequence is repeated while iteratively increasing the pulse length $\tau_i$, resulting in Rabi oscillations after demodulation with the reference frequency.\\

A typical Rabi oscillation for varying microwave powers is depicted in Fig.\,\fref{fig:S8}{a}. The microwave frequency was set to \SI{2.87}{GHz}, which results in a strong damping of the oscillation due to a beating of the different NV resonances, the varying Rabi frequencies throughout the thick diamond sample and the influence of local electric fields \cite{jamonneau2016electricnoise}. We decided not to split up the resonances by applying an external magnetic bias-field as the microwave-induced cavity signal changes in that case were too small. Consequently, we performed all measurements in a magnetic zero-field configuration enabling an increased NV signal when overlapping all the NV resonances. For all microwave powers, the maximum of the damped oscillation was fitted with a parabola (green) to obtain the $\pi$-pulse duration $t_{\pi}$ yielding a Rabi frequency of $\Omega=\pi/t_{\pi}$. We measured the Rabi frequencies for a set of microwave and pump powers portrayed in Fig.\,\fref{fig:S8}{b}. For increasing microwave power $P_{\text{MW}}$ the Rabi frequency $\Omega$ is increased due a stronger magnetic field $B_{1}$ driving the spin transition, $\Omega\propto B_1\propto\sqrt{P_{\text{MW}}}$ \cite{barry2020review}. The weak pump power dependence of the Rabi frequency can be attributed to a repolarization effect during the microwave pulses because of continuously optically pumping the NV centers. With the shown combinations of microwave- and pump powers, individual singlet ACSs were determined (see Fig.$\,$\ref{fig:S9}) from which we calculated the average result portrayed in Fig.$\,$5 in the main text. 

\subsection{Individual absorption cross sections}
\begin{figure*}[tb]
    \centering
    \includegraphics[]{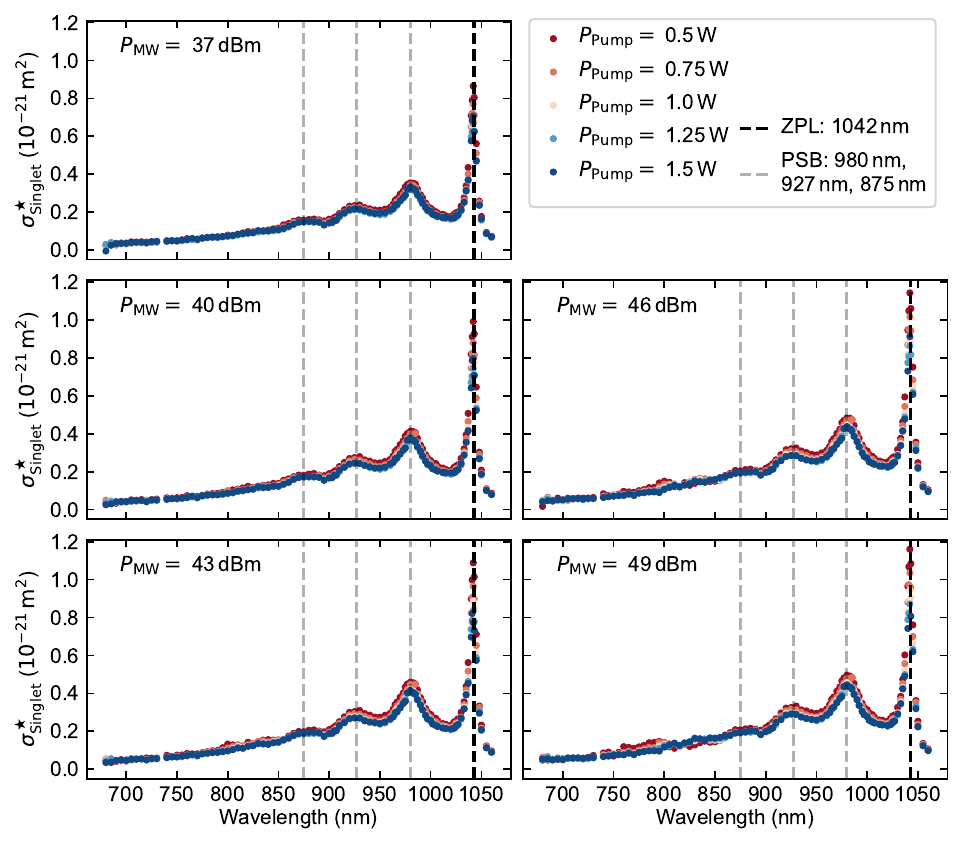}
    \caption{\textbf{Singlet absorption cross sections obtained for different pairs of pump and microwave powers.} We used pump powers ranging from \SI{0.5}{W} to \SI{1.5}{W} and microwave powers ranging from \SI{37}{dBm} to \SI{49}{dBm}. Each pair of pump and microwave powers yields an ACS over the entire spectral range using a different microwave-induced absorption coefficient $\mu$ and number of additional singlet absorbers $\Delta N_{\text{Singlet}}$. All results are in good agreement with each other, thereby cross-validating our experimental method. The ACS shown in Fig.$\,$5 in the main text is the average over all of these individual ACSs.}
    \label{fig:S9}
\end{figure*}

To validate our employed method to determine the singlet ${}^1E\rightarrow{}^1A_1$ ACS, we used the experimental approach described above for various combinations of pump and microwave powers. Each set of pump and microwave powers leads to different amplitudes of the cavity resonances $A_{\text{Pump}}$ and $A_{\text{MW}}$. As a result, each set of powers yields a different absorption coefficient according to Eq.$\,$(\ref{eq:mu}). Furthermore, each set of powers induces a different singlet population which we map with the help of the employed rate model calculations. In this model, the excitation rate $\Lambda_{\text{NV}}$ and Rabi frequency $\Omega$ change with each set of powers, leading to a distinct $\Delta N_{\text{Singlet}}$ which describes the number of additional singlet absorbers due to the resonant microwave field which correspond to $\mu$. All the individual ACSs resulting from these pairs of powers are presented in Fig.$\,$\ref{fig:S9} and agree well with each other. To capture the slight deviations depending on the exact pump and microwave power within our final result, we use the average of all individual ACSs, which we present with the corresponding standard deviation in Fig.$\,$5 in the main text. In Fig.\,\ref{fig:S9}, the \SI{532}{nm} absorption cross section $\sigma^{\,\bigstar}_{532}$ \cite{abs532_0_95} was used. For the alternative $\sigma^{\,\blacktriangle}_{532}$ \cite{wee2007sigma532_0_31}, the individual singlet ACSs agree just as well with each other.

\bibliography{bibliography}
\end{document}